\documentclass[preprints,article,accept,moreauthors,10pt,a4paper]{mdpi}

\usepackage{graphicx}

\newcommand{\keV}{\rm{\, keV }}

\newcommand{\beq}{\begin{equation}}
\newcommand{\eeq}{\end{equation}}
\newcommand{\ba}{\begin{array}}
\newcommand{\ea}{\end{array}}

\newcommand{\pre}   {{\it Physical Review E}}

\newcommand{\apj}{\mbox{\it Astrophys. J.}}
\newcommand{\apjl}{\mbox{\it Astrophys. J.}}

\newcommand{\apss}{\mbox{\it Astrophys. ans Space Science}}
\newcommand{\aap}{\mbox{\it Astron. Astrophys.}}

\newcommand{\araa}{\mbox{\it Annu. Rev. Astron. Astrophys.}}

\newcommand{\jcap}{\mbox{\it J. Cosmol. Astropart. Phys.}}
\newcommand{\mnras}{\mbox{\it Mon. Not. R. Astron. Soc.}}
\newcommand{\nar}{\mbox{\it New Astronomy Rev.}}
\newcommand{\nat}{\mbox{\it Nature}}
\newcommand{\na}{\mbox{\it New A.}}

\newcommand{\physrep}{\mbox{\it Phys. Rep.}}

\newcommand{\planss}{\mbox{\it Planetary and Space Science}}
\newcommand{\ssr}{\mbox{\it Space Science Rev.}}
\def\lsim{\raisebox{-0.3ex}{\mbox{$\stackrel{<}{_\sim} \,$}}}
\def\gsim{\raisebox{-0.3ex}{\mbox{$\stackrel{>}{_\sim} \,$}}}

\firstpage{1}
\pubvolume{xx}
\articlenumber{1}
\pubyear{2018}
\copyrightyear{2018}
\externaleditor{Academic Editor: name}
\history{Received: date; Accepted: date; Published: date}

\Title{Plasmas in Gamma-Ray Bursts: particle acceleration, magnetic fields, radiative Processes and environments}

\Author{Asaf Pe'er}

\AuthorNames{Asaf Pe'er}

\address{%
$^{1}$ \quad Physics Department, Bar Ilan University, Ramat-Gan 52900, Israel; asaf.peer@bia.ac.il; Tel.: +972-3-5318438\\
}

\abstract{Being the most extreme explosions in the universe, gamma-ray bursts
(GRBs) provide a unique laboratory to study various plasma physics
phenomena. The complex lightcurve and broad-band, non-thermal spectra
indicate a very complicated system on the one hand, but on the other
hand provide a wealth of information to study it. In this chapter I
focus on recent progress in some of the key unsolved physical
problems. These include: (1) Particle acceleration and magnetic field
generation in shock waves; (2) Possible role of strong magnetic fields
in accelerating the plasmas, and accelerating particles via magnetic
reconnection process; (3) Various radiative processes that shape the
observed lightcurve and spectra, both during the prompt and the
afterglow phases, and finally (4) GRB environments and their possible
observational signature.}

\keyword{acceleration of particles; gamma-ray bursts; jets; magnetohydrodynamics; plasmas; radiation mechanism: non-thermal }

\begin{document}


\section{Introduction}

Gamma-ray bursts (GRBs) are the most extreme explosions known since
the big bang, releasing as much as $10^{55}$~erg (isotropically
equivalent) in few seconds, in the form of gamma-rays
\citep{Abdo+09b}. Such a huge amount of energy released in such a
short time must be accompanied by a relativistic motion of a
relativistically expanding plasma. There are two separate arguments
for that. First, the existence of photons at energies $\gsim$~MeV as
are observed in many GRBs necessitates the production of $e^\pm$ pairs
by photon-photon interactions, as long as the optical depth for such
interactions is greater than unity. Indeed, the huge luminosity
combined with small system size, as is inferred from light-crossing
time arguments ensures that this is indeed the case. Second, as has
long been suspected and is well established today, small baryon
contamination, originating from the progenitor - being either a
single, collapsing star or the merger of binary, degenerate stars
(e.g., neutron stars or white dwarfs) implies that some baryon
contamination is unavoidable. These baryons must be accelerated for
the least by the radiative pressure into relativistic velocities.

This general picture was confirmed already 20 years ago by the
detection of afterglow - a continuing radiation that is observed at
late times, up to weeks, months and even years after the main GRB, at
gradually lower frequencies - from X-ray to radio \citep{Costa+97,
  VanPar+97, Frail+97}. Lasting for many orders of magnitude longer
than the prompt phase, this afterglow radiation is much easier to
study. Indeed, it had been extensively studied in the past two
decades, after the initial detection enabled by the Dutch-Italian
Beppo-SAX satellite.

\begin{figure}
\includegraphics[width=12cm]{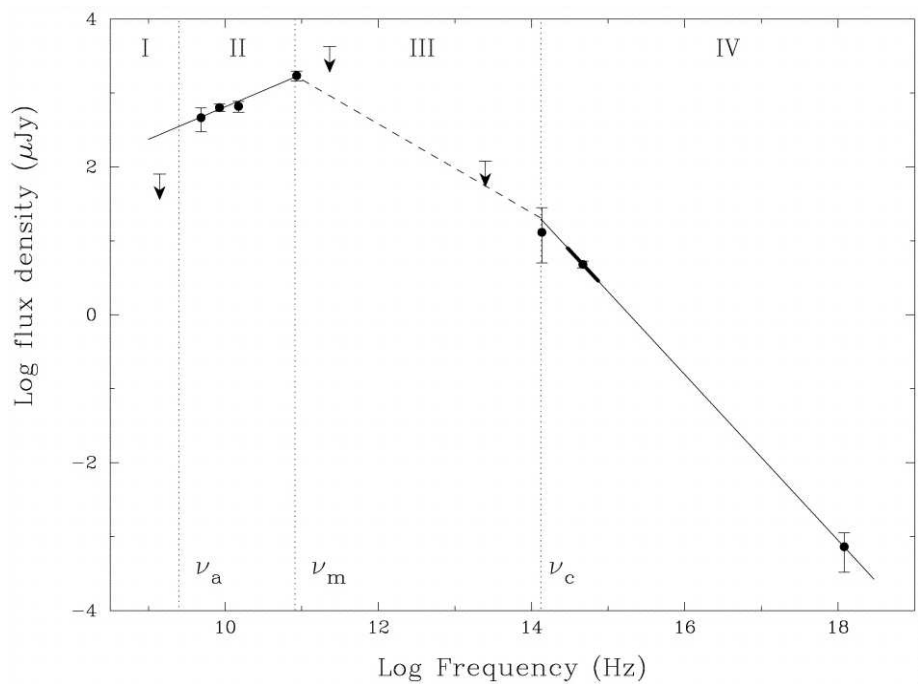}
\caption{ X-ray to radio spectrum of GRB970508 taken 12 days after
  the event is well fitted by a broken power law, as is expected from
  a power law distribution of electrons that emit synchrotron
  radiation. Marked are the transition frequencies: the self
  absorption frequency $\nu_a$, the peak frequency $\nu_m$ and the
  cooling frequency $\nu_c$. This figure is taken from
  \citet{Galama+98a}.}
\label{fig:1}
\end{figure}

Fitting the observed spectra shows a clear deviation from a black-body
spectra. Instead, the afterglow of many bursts is well fitted by
synchrotron radiation from a power law distribution of radiative
electrons \citep{WRM97, Galama+98, WG99}; see Figure \ref{fig:1}. The late time decay is well explained
by the gradual velocity decay of the expanding plasma as it propagates
into the surrounding medium. This decay is well-fitted (at late times)
by the Blandford-Mckee self-similar solution \citep{BM76}  of a
relativistic explosion (I ignore here the early afterglow phase which
typically lasts a few minutes, as during this phase the decay does not
follow a simple self-similar law, and is as of yet not fully
understood).

At the onset of the afterglow phase, the velocity of the expanding plasma is very close to the speed of
light, with initial Lorentz factor of few hundreds. This is greater
than the speed of sound, $c/ \sqrt{3}$, and as such necessitates the
existence of a highly relativistic shock wave. The combined temporal
and spectral analysis thus led to the realization that, at least
during the afterglow phase, a relativistic shock wave must exist. This
shock wave expands into the circumburst medium, and gradually slows
down as it collects and heats material from it.

Interpreting the observed signal during the afterglow phase in the
framework of the synchrotron emission model, one finds that the
inferred values of the magnetic fields, $\lsim 1$~G, are about two
(and in some cases more) orders of magnitude stronger than the
compressed values of the circumburst magnetic field \citep{WRM97,
  KuPa00, SBK14}. This implies that in order to explain the observed
signal, the relativistic shock wave must be able to both (1)
accelerate particles to high energies, producing a non-thermal
(non-Maxwellian) distribution of particles; and (2) generate strong
magnetic field, which causes the energetic particles to radiate their
energy via synchrotron emission.

Studies of the afterglow phase by themselves therefore lead to several
very interesting plasma physics phenomena which are not well
understood, and are at the forefront of current research. These
include (1) the physics of relativistic shock waves, both propagation
and stability; (2) particle acceleration to non-thermal distributions;
(3) generation of strong magnetic fields; and (4) radiative processes
that lead to the observed spectra.

Still, the prompt phase of GRB is considered even more challenging. As
its name implies, this phenomenon lasts only a short duration of time,
typically a few seconds. As opposed to the afterglow phase, this stage
is characterized by fluctuative, non-repeating lightcurve, with no two
GRB lightcurves similar to each other. Furthermore, its spectra does
not resemble neither a black body (Planck) spectrum, nor- as has been
realized in the past decade - that of a synchrotron emission from a
power law distribution of electrons, as in the afterglow phase.
Though the large diversity within the bursts prevented, so far, clear
conclusions.

Another very challenging aspect is that the origin of the rapid acceleration that results in the relativistic expansion is not
yet fully understood. While it was initially thought to occur as a
result of the photon's strong radiative pressure (the ``fireball''
model), in recent years it is argued that strong magnetic fields -
whose origin may be associated with the progenitor(s), hence external
to the outflow, may play a key role in the acceleration process. If
this is indeed the case, the plasma must be magnetically dominated,
namely $u_B \gg \{u_k, u_{th}\}$, where $u_B, u_k$ and $u_{th}$ are the
magnetic, kinetic and thermal energies of the plasma, respectively.

Either way, the plasma in GRBs during its prompt emission phase is
characterized by strong interactions accompanied by energy and momentum
exchange between the particle and photon fields, and / or the
particles and the magnetic fields. Combined with the different
conditions during the afterglow phase, one can conclude that GRBs
provide a unique laboratory to study various fundamental questions in
plasma physics. These are related to the creation of magnetic fields,
acceleration of particles, emission of radiation and the interaction
between all these three fields. Furthermore, the relativistic
expansion can lead to the developments of several instabilities in
the expanding plasma, which, in turn, can affect the phenomena
previously mentioned.

In this chapter, I highlight the current state of knowledge in these
areas. I should stress that I limit the discussion here to plasma
physics phenomena only; in recent years, there have been many
excellent reviews covering various aspects of GRB phenomenology and
physics, and I refer to reader to these reviews for a more
comprehensive discussion on the various subjects. A partial list
includes \citet{AB11, GM12, Bucciantini12, GR13, Daigne13, Zhang14,
  Berger14, MR14, Peer15, KZ15, Granot+15, Zhang+16, Toma+16, PR17,
  BM17, DDM17, VE18, Nagataki18}.

Of the many plasma physics effects that exist is GRBs - some of
them unique to these objects, I discuss here several fundamental
phenomena which emerge directly from GRB studies. Due to the wealth of
the subject, I can only discuss each topic briefly. In each section, I
refer the reader to (some) relevant literature for further discussion.
The topics I cover here include the following. Acceleration of
particles by relativistic shock waves are discussed in section
\ref{sec:2}. Section \ref{sec:3} is devoted to magnetic fields. I
discuss generation of magnetic fields by shock waves in section
\ref{sec:3.1}, and their possible role during the prompt emission
phase in section \ref{sec:3.2}. I briefly discuss the acceleration of
particles in magnetically dominated outflow via reconnection of
magnetic field lines in section \ref{sec:3.3}.  I then discuss the
radiation field, which play a key role in GRBs in section
\ref{sec:4}. I first introduce the ``classical'' radiative processes
in section \ref{sec:4.1} and then introduce the photospheric emission
in section \ref{sec:phot_emission}.  Finally, I very briefly consider
the different environments into which GRBs may explode and their
effects in section \ref{sec:5}, before concluding.

\section{Acceleration of particles in shock waves}
\label{sec:2}

The idea that shock waves can be the acceleration sites of particles
dates back to Enrico Fermi himself \citep{Fermi49, Fermi54}, and had
been extensively studied over the years since \citep{ALS77, BO78, Bell78,
  BE87, JE91, MD01, Bell04}. The key motivation was to explain the observed
spectrum and flux of cosmic rays. Fermi's original idea suggests that
particles are energized as they bounce back and forth across the shock
wave. Its basic details can be found today in many textbooks \citep[e.g.,][]{Longair11}.

In the context of GRBs, it was proposed at the mid 1990's that the relativistic shock
waves that exist in GRB plasmas may provide the conditions required
for the acceleration of particles to the highest observed energies,
$\gsim 10^{20}$~eV \citep{MU95, W95, Vietri95}. While this idea is
still debatable \citep[e.g.,][]{Filip18}, the observations of $>$~GeV,
and up to $\sim 100$~GeV photons \citep{Ackermann+14} during the
prompt phase of several GRBs implies that very high energy particles
must exist in the emitting region. While these particles can be
protons, energetically it is much less demanding if these are
electrons that are accelerated to non-thermal distribution at high
energies. This is due to the lighter mass of the electrons, which
implies much more efficient coupling to the magnetic and photon
fields, hence much better radiative efficiency. These energetic
particles, in turn, radiate their energy in the strong magnetic fields
that are believed to exist, as well as Compton scatter the photons to
produce the very high energy photons observed.

Fitting the observed spectra of many GRBs in the framework of the
synchrotron model (namely, under the assumption that the leading
radiative mechanism is synchrotron emission by energetic electrons)
strongly suggests that the radiating particles do not follow a
(relativistic) Maxwellian distribution. Rather, they follow a power
law distribution at high energies, $dn_E/dE \propto E^{-p}$, with
power law index $p \approx 2.0 - 2.4$ \citep{Tavani96, WRM97,
  Crider+97, Gal+98, Berger+03a}. This power law distribution is
exactly what is expected from acceleration of particles in shock waves
within the framework of the Fermi mechanism \citep[e.g.,][]{BO78,
  Bell78, BE87, SKL15}. Intuitively, the power law shape of the
distribution can be understood as there is no characteristic momentum
scale that exists during the acceleration process, implying that the
rate of momentum gain is proportional to the particle's momentum.

The power law index inferred from observations is close to Fermi's
original suggestion of $2.0$. This is surprising, given that the shock
waves in GRBs both during the prompt (if exist) and afterglow phases
must be relativistic, while Fermi's work dealt with ideal,
non-relativistic shocks.

In fact, the situation is far more complicated.  Despite many decades
of research, Fermi process is still not fully understood from first
principles. This is attributed mainly to the highly non-linear
coupling between the accelerated particles and the turbulentic
magnetic field at the shock front. The magnetic field is both
generated by the energetic particles (via the generated currents) and
at the same time affects their acceleration. This makes analytical
models to be extremely limited in their ability to simultaneously
track particle acceleration and magnetic field generation.

Due to this complexity, most analytical and Monte-Carlo methods use
the ``test particle'' approximation. According to this approximation,
during the acceleration process the accelerated particles interact
with a fixed background magnetic field. These models therefore neglect
the contribution of the high energy particles to the magnetic field,
which occurs due to the currents they generate. This assumption can be
justified as long as the accelerated particles carry only a small
fraction of the available energy that can be deposited to the magnetic
field. However, as explained above, this assumption is {\bf not}
supported by current observations.

Furthermore, relativistic shocks, as are expected in GRBs, introduce
several challenges which do not exist when considering
non-relativistic shocks. These include (1) the fact that the
distribution of the accelerated particles cannot be considered
isotropic; (2) mixing of the electric and magnetic fields when moving
between the upstream and downstream shock regions; and (3) the fact
that it is more difficult to test the theory (or parts of it), and one
has to rely on very limited data, which can often be interpreted in
more than a single way.

Very broadly speaking, theoretical works can be divided into three
categories. The first is semi-analytic approach \citep[e.g.,][]{KH89,
  Malkov97, Kirk+00, CAP10, KW05, Keshet17}, in which particles are
described in terms of distribution functions, enabling analytic or
numerical solutions of the transport equations. Clearly, while this is
the fastest method, reliable solutions exist only over a very limited
parameter space region, and several considerable simplifications (e.g.,
about the turbulence, anisotropy, etc.) are needed. The second method
is the Monte-Carlo approach \citep{KS87, Ellison+90, Achterberg+01,
  LP03, ED04, SB12, Ellison+13, Bykov+17}. In this method, the
trajectories and properties of representative particles are tracked,
assuming some average background magnetic fields. The advantage of
this method is that it enables to explore much larger parameter space
region than analytical methods while maintaining fast computational
speed. The disadvantages are (a) the simplified treatment of the
background magnetic field, which effectively implies that the ``test
particle'' approximation is used; and (b) current Monte-Carlo codes
use a simplified model to describe the details of the interactions
between the particles and magnetic fields. For example, many codes use
the ``Bohm'' diffusion model, which is not well-supported
theoretically \citep[see][]{RP18}.

The third approach is the Particle-In-Cell (PIC) simulations
\citep{Silva+03, Frederiksen+04, Spit08, SS09, Nishikawa+16}. These
codes basically solve simultaneously both particle trajectories and
electromagnetic fields in a fully self-consistent way. They therefore
provide the ``ultimate answer'', namely the entire spectra of the
accelerated particles alongside the generated magnetic field. They
further provide the details of the generated magnetic turbulence as
well as visualize the formation process of collisionless
shocks. However, these codes are prohibitively computationally
expensive, and are therefore limited to a very small range both in
time and space. Modern simulations can compute processes on a length scale of no more than a few
thousands of skin depth ($c/\omega_p$). This scale is many orders
of magnitudes - typically 7 - 8 orders of magnitude shorter than the
physical length scale of the acceleration region, as is inferred from
observations. This is an inherent drawback, that cannot be overcome
in the nearby future.

To conclude this section, GRB observations provide a direct evidence -
possibly the most detailed observational evidence for the acceleration
of particles in relativistic shock waves. These evidence triggered a
huge amount of theoretical work aimed at understanding this phenomenon
from first physical principles. Due to its huge complexity, and while
a huge progress was made in the past two decades or so mainly due to
advance in PIC codes, the problem is still far from being solved.

\section{Magnetic fields in GRBs}
\label{sec:3}

In addition to the existence of clear evidence that shock waves in GRBs serve as particle
acceleration sites, there is a wealth of evidence for the existence of
strong magnetic fields in GRBs.  When discussing magnetic fields in
GRBs, one has to discriminate between two, very different scenarios.

First, as already discussed above, fitting the data of GRB afterglow
strongly suggests that the main radiative mechanism during this phase is
synchrotron emission from energetic electrons. This idea therefore
implies that strong magnetic fields must exist in the plasma. Fitting
the afterglow provide evidence that the magnetic fields are about two
orders of magnitude - and in some cases more - than the values
expected from compression of the intergalactic field \citep{WRM97,
  KuPa00, Yost+03, SBK14, GFP18}. This provides an indirect evidence
that the relativistic shock wave that inevitably exists during this
phase must generate strong magnetic field, in parallel to accelerating
particles.

Second, while no direct evidence currently exists, it had been
proposed that during the prompt emission phase, the GRB plasma may in
fact be Poynting-flux dominated \citep{SDD01, Drenkhahn02, DS02, VK03,
  Giannios05, Giannios06, GS05, MR11}. If this idea is correct, then
the origin of the magnetic field must be external to the plasma -
namely originate at the progenitor. In this scenario, the magnetic
field serves as an energy reservoir that is used to both accelerate
the plasma and at the same time accelerate particles to high energies.

\subsection{Magnetic field generation in shock waves}
\label{sec:3.1}

As is typical to most (in fact, nearly all) astrophysical plasmas, and
certainly in GRBs, the shock waves that exist are collisionless,
namely they are not mediated by direct collisions between the particles
(as opposed, to, e.g., shock waves that occur while a jet plane
exceeds the speed of sound in the earth's atmosphere). This can easily
be verified for the shock wave in the afterglow phase of GRBs by
considering the mean free path for particle interaction, $l = (n
\sigma_T)^{-1} \simeq 10^{24}$~cm. Here, $n \simeq 1~{\rm cm^{-3}}$ is
the typical interstellar medium (ISM) density, and $\sigma_T$ is
Thompson cross section, which is the typical cross section for
particle interaction. This scale is many orders of magnitude longer
than the scale of the system, implying that the generated shock waves
must be collisionless.

Instead of direct collisions, the shock waves are generated by
collective plasma effects, namely the charged particles generate
currents. These currents, in turn generate magnetic fields that
deflect the charged particles trajectories, mixing and randomizing
their trajectories, until they isotropize. Thus, the generation of
collisionless shock waves must include generation of turbulentic
magnetic fields. The key questions are therefore related to the details
of the process, which are not fully understood. These include: (1) the
nature of the instability that generates the turbulentic field; (2)
the strength - and scale of the generated field; and (3) the
interconnection between the particle acceleration process and the
magnetic field generation process.

\begin{figure}
\includegraphics[width=12cm]{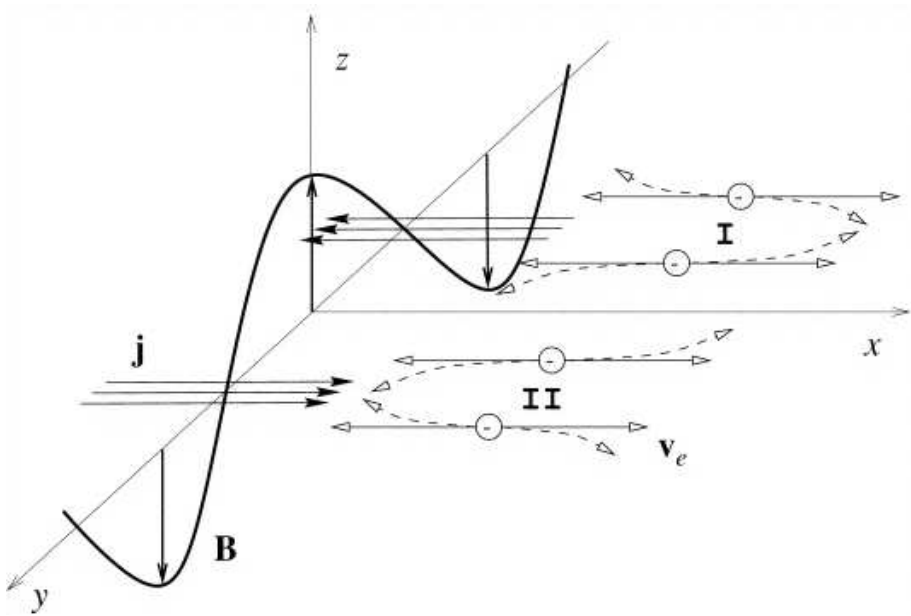}
\caption{ Illustration of the Weibel (also denoted "relativistic two stream") instability, taken from \citet{ML99}. A magnetic field perturbation deflects electron motion along the x-axis, and results in current sheets (j) of opposite signs in regions I and II, which in turn amplify the perturbation. The amplified field lies in the plane perpendicular to the original electron motion.  }
\label{fig:2}
\end{figure}

The most widely discussed mechanism by which (weakly magnetized)
magnetic fields can be generated is the Weibel instability
\citep[e.g.,][]{Weibel59, ML99, GW99, WA04, LE06, AW07, SLE12, KEG15,
  Bret+16, Pelletier+17, BP18}. In this model, small fluctuation in
the magnetic fields charge separation opposite charges in the
background plasma. These particles then form "filaments" of
alternating polarity, which grow with time, as the currents carried by
the charged particles positively feed the magnetic fields. This is illustrated in Figure \ref{fig:2}, taken from \citet{ML99}.  Indeed,
this instability is routinely observed in many PIC simulations
\citep{Silva+03, Frederiksen+04, Nishikawa+05, Kato05, RRNH07,
  Spit08a, Spit08, Nishikawa+09}, which enable to quantify
it. Furthermore, these simulations prove the ultimate connection
between the formation of collisionless shock waves and the generation
of magnetic fields.

\begin{figure}
\includegraphics[width=12cm]{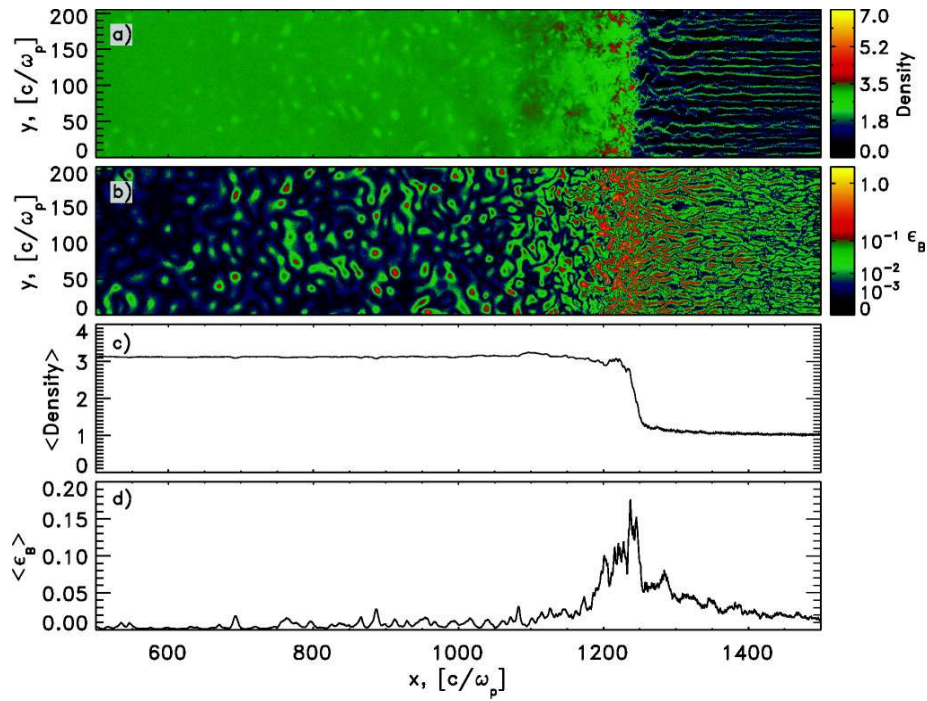}
\caption{ Snapshot of a region from a large 2D relativistic PIC shock simulation, taken from \citet{Chang+08}.
(a) Density structure in the simulation plane showing the plasma density enhancements in the foreshock region that steadily grow up to the shock transition region, where the density becomes homogeneous. (b) Magnetic energy, normalized in terms of upstream energy of the incoming flow.  The upstream magnetic filaments, which can be visualized as sheets coming out of the page, that are formed by the Weibel instability reach a peak just before the shock. (c) Plasma density averaged in the transverse direction as a function of the distance along the flow. (d) Magnetic energy density averaged in the transverse direction, as a function of distance along the flow. Clearly, strong magnetic fields are generated, but quickly decay.
  }
\label{fig:3}
\end{figure}

However, these same simulations show that the generated magnetic
fields decay over a relatively very short length scale, of few tens -
few hundreds of plasma skin depth, as was suggested earlier
\citep{Gruzinov01, Spit05, MN06a, Chang+08}; see Figure \ref{fig:3}. This is in sharp contrast
to the observed synchrotron signal, which requires that the magnetic
field, necessary for the synchrotron emission will remain substantial
over a much larger scale, comparable to the scale of the system.

This drawback, clearly observed in modern PIC simulations, triggered a
few alternative suggestions. First, it was suggested that the prompt
emission can possibly be generated over a much shorter scale than
previously thought \citep{PZ06}. Other works investigate the effect of
energetic particles (resulting from the acceleration process) on the
evolution of the magnetic fields. It was argued \citep{MN06a, MN06b,
  Couch+08} that strong magnetic fields can last over a substantial
range due to other types of instabilities.  It was further suggested that the gradual increase in the
population of high energy particles that results from the Fermi
acceleration process, gradually increases the characteristic length
scale of the magnetic field \citep{Keshet+09}.  Other suggestions
include macroscopic turbulence that is generated by larger scale
instabilities that take place as the shock waves propagate through a
non homogeneous media. Indeed, inevitable density fluctuations in the
ambient medium will trigger several instabilities (e.g., Richtmyer -
Meshkov or kink) that can in principle grow over a large scale
\citep{SG07, ZMW09, Inoue+11, Mizuno+11, Mizuno+14}. Another possibility, which is very realistic in GRB environment, is generation of mgnetic fields by various instabilities (such as kinetic Kelvin-Helmholtz, mushroom or kink instability) that are stimulated if the relativistic jet is propagating into an already magnetized plasma \citep{Mizuno+14a, Nishikawa+17}. Indeed, helical magnetic fields may be important in jet acceleration and collimation (see the following section), and their existence will stimulate turbulence as the jet propagates through the plasma. This scenario differs than that presented in Figure \ref{fig:3}, as it includes both reverse and forward shocks, as well as contact discontinuity \citep{Ardaneh+16}, all provide possible sites for enhancement of magnetic turbulence.

Thus, over all, the origin of the magnetic field as is required to
produce the observed (synchrotron) radiation is still an open
question. This field remains one of the very active research fields.

\subsection{Highly magnetized plasma during GRB prompt emission?}
\label{sec:3.2}

\subsubsection{Motivation}

Very early on it was realized that the extreme luminosity, rapid
variability and $>$MeV photon energies imply that GRB plasma must be
moving relativistically during the prompt emission phase, otherwise
the huge optical depth to pair production, $\tau \gsim 10^{15}$ would
prevent observation of any signal \citep[see, e.g,][for
  reviews]{Piran04, Meszaros06, Peer15}. This idea was confirmed by
the late 1990's with the discovery of the afterglow, that proved that
the plasma indeed propagates at relativistic speeds.

Thus, {\bf two major episodes of energy conversion exist}: first,
the conversion of the gravitational energy to kinetic energy - namely,
the acceleration of plasma that results in the generation of
relativistic jet. Second, the huge luminosity suggests that a
substantial part of the kinetic energy is dissipated, and used to heat
the particles and generate the observed signal.

Originally, it was argued that instabilities within the expanding
plasma would generate shock waves, which are internal to the flow
\citep[``internal shocks''; see][]{RM94, SP97, RF00}.  By analogy with
the afterglow phase, it was then suggested that a very similar
mechanism operates during the prompt phase. Shock waves generated by
internal instabilities both generate strong
magnetic fields and accelerate particles, which in turn emit the observed prompt radiation
\citep{RM94, Kobayashi+97}. In the framework of this model, the
internal shocks are therefore the main mechanism of kinetic energy
dissipation, and magnetic fields "only" provide the necessary
conditions needed for synchrotron radiation.

While this scenario gained popularity by the late 1990's, it was soon
realized that it suffers several notable drawbacks. First, the very
low efficiency in kinetic energy dissipation, of typically a few \%
\citep{MMM95, PSM99, Beloborodov00, Spada+00, GSW01, PLC17}. This can
be understood, as only the differential kinetic energy between the
propagating shells can be dissipated by internal collisions. The only
way to overcome this problem is by assuming a very high contrast in
the Lorentz factors of the colliding shells  \citep{KS01}.

Second, once enough data became available, it became clear that as
opposed to the afterglow phase, the simplified version of the
synchrotron model does not provide acceptable fits to the vast
majority of GRB prompt emission spectra \citep{Crider+97, Preece+98,
  Preece+02, GCG03, AB15}. Thus, one has to consider alternative
emission mechanisms, or, for the least, consider ways to modify the
synchrotron emission model (see further discussion in section
\ref{sec:4} below).

Third, the details of the initial explosion that triggered the GRB and
the mechanism that produce relativistic motion (jet) in the first
place remain uncertain.  One leading model is the "Collapsar" model
\citep{Woosley93, MW99}, according to which the core collapse of a
massive star triggers the GRB event. In this scenario, the main energy
mediators are neutrinos that are copiously produced during the
collapse, and transfer the gravitational energy to the outer stellar
regions, which are accelerated to relativistic velocities.

An alternative model for the formation of relativistic jets is the
mechanism first proposed by \citet{BZ77}. According to this idea,
rotational energy and angular momentum are extracted from the created
rapidly spinning (Kerr) black hole by strong currents. In this scenario, strong magnetic
fields play a key role in the energy extraction process. Thus, the
emerging plasma must be Poynting-flux dominated, and the kinetic
energy is sub-dominant.

This idea has two great advantages. First,
the rotation of a rapidly spinning black hole provides a huge energy
reservoir that can in principle be extracted. Second, this mechanism
is fairly well understood, and is believed to exist in
nature. Furthermore, it does not suffer from the low efficiency
problem of the ``internal shock'' scenario. Indeed, this mechanism
gained popularity over the years, and is in wide use for explaining
energy extraction in other astronomical objects, such as active
galactic nuclei \citep[AGNs; see][]{BBR84, WC95} or X-ray binaries
\citep[XRBs;][]{Narayan+07}.

I should stress that as of today, there is no clear evidence that points to which
of the two scenarios act in nature to produce the relativistic GRB
jets - or possibly a third, as of yet unknown scenario. However, the
possibility that strong magnetic fields may exist motivated studies of
the dynamics of highly magnetized plasmas. Under this hypothesis of
Poynting-dominated flow, one need to address two independent
questions. The first is the creation of the relativistic jet (namely,
the acceleration of the bulk outflow to highly relativistic
velocities). The second is the acceleration of (individual) particles
to high energies, needed to explain the observed radiative signal.

\subsubsection{Detailed models}

As opposed to the ``internal shock'' model, the basic idea in the
``Poynting-flux dominated models'' is that the strong magnetic fields
serve as ``energy reservoir''. The magnetic energy is converted to
kinetic energy and heat (or particle acceleration) by reconnection of
the magnetic field lines. In the past few years, many authors
considered this possibility. Very crudely speaking , one can divide
the models into two categories. The first assumes continuous magnetic
energy dissipation \citep[e.g.,][]{Usov92, Thompson94, VK01,
  Drenkhahn02, DS02, VK03, LB03, Levinson06a, Giannios08,
  Komissarov+09, BG17}. These models vary by the different assumptions about
the unknown rate of reconnection, outflow parameters, etc.  The second
type assumes that the magnetic dissipation - hence the acceleration
occurs over a finite, short duration [\cite{Cont95, Tchekhovskoy+08,
    TNM10b, Komissarov+10, Metzger+11, Granot+11, ZY11,
    Sironi+15}]. The basic idea is that variability in the central
engine leads to the ejection of magnetized plasma shells, that expand
due to internal magnetic pressure gradient once they lose causal
contact with the source.

\begin{figure}
\includegraphics[width=12cm]{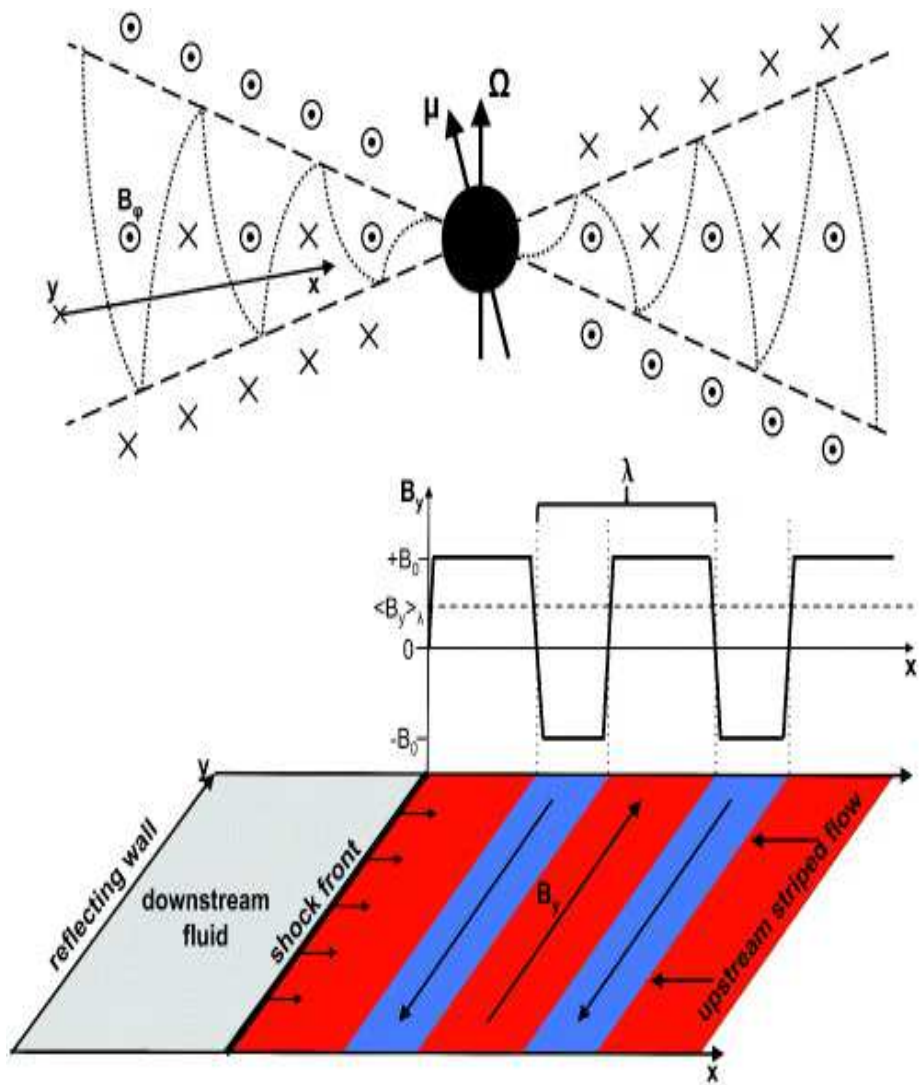}
\caption{ Upper panel: poloidal structure of the striped pulsar wind, according to the solution of \citet{Bogovalov99}. The arrows denote the pulsar rotational axis (along $\Omega$, vertical) and magnetic axis (along $\mu$, inclined). Within the equatorial wedge bounded by the dashed lines, the wind consists of toroidal stripes of alternating polarity, separated by current sheets (dotted lines). Lower panel: 2D PIC simulation setup geometry. Figure is taken from \citet{SS12}.
  }
\label{fig:4}
\end{figure}

A relatively well understood scenario is the ``striped wind'' model,
first proposed in the context of pulsars \citep{KC84,
  Coroniti90}. According to this model, the gravitational collapse
that triggers the GRB event leads to a rapidly rotating, highly
magnetized neutron star (which can later collapse into a black
hole). The rotational axis is misaligned with its dipolar
moment, which naturally produces a striped wind above the
light-cylinder (see Figure \ref{fig:4}). This striped wind consists of cold regions with
alternating magnetic fields, separated by hot current
sheets. Reconnection of magnetic field lines with opposite polarity is
therefore a natural consequence; such reconnection leads to the
acceleration of the wind.

Evolution of the hydrodynamic quantities in these Poynting-flux
dominated outflow within the ``striped wind'' model was considered by
several authors \citep{LK01, SDD01, Drenkhahn02, DS02, VK03, SD04,
  Giannios05, Giannios06, GS05, MR11, BPL17}.  The scaling laws of the
acceleration can be derived under the ideal MHD limit approximation,
which is a good approximation due to the high baryon load
\citep{SDD01}. Furthermore, in this model, throughout most of the jet
evolution the dominated component of the magnetic field is the
toroidal component, and so the magnetic field is perpendicular to the
outflow direction, $\vec B \perp \vec \beta$. Under these assumptions,
it can be shown that the standard equations of mass, energy and
momentum flux conservations combined with the assumption of constant
reconnection rate (which is not specified in this model) leads to a
well defined scaling law of the Lorentz factor, $\Gamma (r) \propto
r^{1/3}$. This is different than the scaling law expected when the
acceleration is mediated by photon field, as originally proposed in
the classical ``fireball'' model, $\Gamma (r) \propto
r$. Furthermore, these scaling laws lead
to testable predictions about the total luminosity that can be
achieved in each of the different phases \citep{Peer17}. However, so far, these were not confronted with observations.

The main uncertainty of these models remains the unknown rate in which
the reconnection process takes place. This rate is model dependent,
and in general depends on the rate of magneto-hydrodynamic (MHD)
instabilities that destroy the regular structure of the flow
\citep{Lyubarsky92, Begelman98, GS06, Gill+18, SL18}. Furthermore,
the presence of strong radiative field can affect this rate
\citep{BPL17}. It should be noted that several PIC simulations predict a nearly universal reconnection rate, of $\sim 0.1 c$ for highly magnetized flows \citep{Riquelme+12, Melzani+14, Guo+15}. This rate is dictated by the dynamics of the plasmoid instability. However, due to the limitations of existing PIC codes, I think it is fair to claim that this is still an open problem.

\subsection{Acceleration of particles in highly magnetized plasma: magnetic reconnection process}
\label{sec:3.3}

While it is natural to envision a highly magnetized progenitor that
results in Poynting-flux dominated outflow in the early stages of GRB
evolution, this possibility leads to two basic questions.  The first
is the details of the reconnection process that dissipates the
magnetic energy. The second is the mechanism by which particles are
accelerated. Observations of non-thermal emission during the prompt
phase necessitates some mechanism that accelerates the particles (for
the least, the electrons) to high energies. However, in
Poynting-dominated flow, this mechanism need to be different than the
celebrated Fermi process. In this environment of highly magnetized
plasmas, both shock formation is limited \citep{Bret+17} and particle
acceleration by shock waves is suppressed \citep{SSA13}.

First, it was shown that the properties of shock waves (if form) and
in particular their ability to accelerate particles to high energies
are different if these shock waves reside in highly magnetized
regions. In this case, the ability of a shock wave to accelerate
particles strongly depends on the inclination angle $\theta$ between
the upstream magnetic field and the shock propagation direction
\citep{SS09b, SS11, Matsumoto+17}. Only if this angle is smaller than a critical
angle $\theta_{crit}$ can the shock accelerate particle
efficiently. At higher angles, charged particles would need to move
along the field faster than the speed of light in order to outrun the
shock, and therefore cannot be accelerated. I point out, though, that these simulations assumed a simple configuration of the initial magnetic field, and thus the results in GRB jets may differ.

\begin{figure}
\includegraphics[width=12cm]{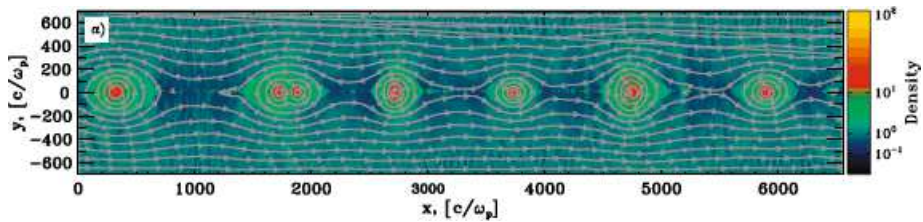}
\caption{Structure of the particle density in the reconnection layer at $\omega_p t=3000$, from a 2D simulation of magnetized plasma having magnetization parameter $\sigma=10$.
Figure is taken from \citet{SS14}.
  }
\label{fig:5}
\end{figure}

On the other hand, particles can be accelerated to high energies by
the reconnection process itself \citep{RL92, Lyutikov03, Jaroschek+04,
  Lyubarsky05, Giannios10b, Lazarian+11, Liu+11, UM11, MU12, BB12,
  Cerutti+12, Cerutti+13b, Kagan+13, US14}; \citep[see][for a recent
  review]{Kagan+15}. The basic idea is that whenever regions of
opposite magnetic polarity are present, Maxwell's equations imply that
there must be a current sheet in between. In this current layer,
magnetic field lines can diffuse across the plasma to reconnect at one
or more ``x''-lines. When particles cross the current sheet, they are
forced back by the reversing magnetic field. This is seen in Figure \ref{fig:5}, taken from \citet{SS14}. The particles can then be
accelerated in the direction perpendicular to the plane of
reconnection by the generated inductive electric fields
\citep{ZH01}. Their energy gain per unit time is therefore $dW/dt = q
E \cdot v \sim q E c$ in the relativistic case.

This general idea had been extensively studied over the years by PIC
simulations, both 2D \citep{ZH01, BB05, ZH07, HZ07, LL08, BB12,
  Cerutti+12, SS14, Guo+14, Nalewko+15, Werner+16, Sironi+16} and 3D
\citep{ZH08, Yin+08, Liu+11, SS11, SS12, Cerutti+14, Guo+15,
  WU17}. These show the generation of hard, non-thermal distribution
of energetic particles that are accelerated at (relativistic)
reconnection sites. These particles follow a power law distribution,
with power law index $p \gsim 2$, for strongly magnetized plasma,
having magnetization parameter $\sigma \equiv u_B/u_{th} \approx 10$
\citep{SS14}. Here, $u_B, u_{th}$ are the magnetic field and thermal
particle energy densities, respectively. While this index is fairly
similar to the one obtained by the Fermi process, it was shown to be
sensitive to the exact value of $\sigma$ \citep{Guo+14, Werner+14}. Early works suggested that in a strongly magnetized plasma, the power law index is $p < 2$, implying that most of the energy is carried by the energetic particles. However, recent simulations that were run for longer times and using larger box sizes, showed convergence towards $p \sim 2$ at late times \citep{PS18}.

A heuristic argument for the power law nature of the particle
distribution was first suggested by \citet{ZH01}, and was demonstrated
by \citet{SS14}. More energetic particles have larger Larmor
radii, and therefore spend more time near the ``x'' point of the
reconnection layer than particles of lower energies. They suffer less
interaction with the reconnected field (in the perpendicular
direction), and therefore spend longer time in the accelerated region,
where strong electric fields exist. Thus, overall, the gained energy
in the acceleration process is proportional to the incoming particle
energy, which results in a power law distribution.

Finally, as discussed in the previous section, jet propagation into an already magnetized plasma triggers and enhances several instabilities, such as kinetic Kelvin-Helmholtz or mushroom instabilities. As was recently shown \citep{Mizuno+14, Nishikawa+17} these instabilities, which have geometries that are different in nature than the "slab" geometry presented in Figures \ref{fig:4}, \ref{fig:5} above, also serve as acceleration sites of particles \citep{Matsumoto+17}.

\section{Photon field in GRB plasmas}
\label{sec:4}

Our entire knowledge (or lack thereof) of GRB physics originates from
the observed electromagnetic signal. As GRBs are the brightest sources
of radiation in the sky, a strong radiation field must exist within
the relativistically expanding plasma. This photon field adds to the
strong non-thermal particle field, and the possible strong magnetic
fields that exist.

Similar to the questions outlined above about the sources and role
played by magnetic fields (being dominant or sub-dominant), one can
divide the basic open questions associated with the photon fields into
two categories. The first is understanding the radiative processes
that lead to the observed signal. The second is to understand the
possible role of the photon field in shaping the dynamics of the GRB outflow.

\subsection{Radiative processes: the classical ideas}
\label{sec:4.1}

The most widely discussed model for explaining GRB emission both
during the prompt and the afterglow phases is synchrotron
emission. This model has several advantages. First, it has been
extensively studied since the 1960's \citep{GS65, BG70} and its
theory is well understood. It is the leading model for interpreting
non-thermal emission in many astronomical objects, such as AGNs and
XRBs.  Second, it is very simple: it requires only two basic
ingredients, namely energetic particles and a strong magnetic
field. Both are believed to be produced in shock waves or magnetic
reconnection process. Third, it is broad band in nature (as opposed,
e.g., to the ``Planck'' spectrum), with a distinctive spectral peak,
that could be associated with the observed peak energy. Fourth, it
provides a very efficient way of energy transfer, as for the typical
parameters, energetic electrons radiate nearly 100\% of their energy
(during the prompt and parts of the afterglow phases).  These
properties made synchrotron emission the most widely discussed
radiative model in the context of GRB emission \citep[e.g.,][ for a
  very partial list]{RM92, MR93, MLR93, MRP94, PX94, PM96, Tavani96,
  Cohen+97, WRM97, SP97b, SPN98, PL98, DM98, WG99, GS02,  Gao+13}.

Synchrotron emission requires a population of energetic electrons.
These electrons, in addition to synchrotron radiation, will inevitably
Compton scatter the emitted photons, producing synchrotron-self Compton
emission (SSC). This phenomenon is expected to produce high energy
photons, that can extend up and beyond the GeV range. Its relative
importance depends on the Compton Y parameter, namely the optical depth
multiplied by the fractional energy change of each photon. This
phenomenon was extensively studied both in the context of the prompt
phase \citep{PW04, BB04, GZ07, AI07} and the afterglow phase in GRBs
\citep{SE01, ZM01a, Harrison+01, GP07, Fan+08, Fraija+12}.  Note that
the results of the scattering does not only affect the photon field
directly, but also indirectly, as the scattering cools the electrons,
hence modified the synchrotron emission. Naturally, the importance of
this non-linear effect depends on the Compton Y parameter, and is
significantly more pronounced during the prompt phase, where the
plasma is much denser and significantly more scattering are expected
than during the afterglow phase.

Observations of high energy photons, above the threshold for pair
creation implies that both pair production and pair annihilation can
in principle take place. If this happens, then a high energy
electromagnetic cascade will occur, namely energetic photons
produce $e^\pm$ pairs, which lose their energy by synchrotron and SSC,
thereby producing another population of energetic photons, etc. These
phenomena further modifies the observed spectra in a non-linear way
\citep{PW04, PW05, GG18}.

A different suggestion is that the main source of emission is not
leptonic, but rather hadronic. This idea lies on the assumptions that
the acceleration process, whose details are of yet uncertain, may be
more efficient in accelerating protons, rather than electrons to high
energies. In this scenario, the main emission mechanism is synchrotron
radiation from the accelerated protons \citep{BD98, Totani98, ZM01a,
  GZ07, Asano+09, Razzaque+10, AM12, CK13}. The main drawback of this
  suggestion is that protons are much less efficient radiators than
  electrons (as the ratio of proton to electron cross section for
  synchrotron emission $\sim (m_e/m_p)^2$). Thus, in order to produce
  the observed luminosity in $\gamma$-rays, the energy content of the
  protons must be very high, with proton luminosity of $\sim 10^{55}-
  10^{56}$~erg~s$^{-1}$. This is at least 3 orders of magnitude higher
  than the requirement from leptonic models.

\subsection{Photospheric emission and GRB dynamics}
\label{sec:phot_emission}

The idea that photospheric (thermal) emission may play a key role as
part of GRB plasma is not new. Already in the very early models of
cosmological GRBs it was realized that the huge energy release, rapid
variability that necessitates small emission radii (due to light
crossing time argument), and the high $\gsim$~MeV photon energy
observed, imply the existence of photon-dominated plasma, namely a
``fireball'' \citep{Pac86, Goodman86, SP90,  Thompson94}.

Initially, therefore, it was expected that the observed GRB spectra
would be thermal. Only with the accumulation of data that showed
non-thermal spectra - both during the prompt and afterglow phases, did
the synchrotron model gain popularity.

While the synchrotron emission model remains the leading radiative
model that can explain the observed signal during the afterglow phase,
it was realized by the late 1990's that it fails to explain the low
energy part of the prompt emission spectra of many GRBs
\citep{Crider+97, Preece+98, Preece+02, GCG03}. Being well understood,
the synchrotron theory provides a robust limit on the maximum low
energy spectral slope that can be achieved. As the observed slope in
many GRBs was found to be harder than the limiting value,
\citet{Preece+98} coined the term ``synchrotron line of
death''. Despite two decades of research, this result is still
debatable \citep{AB15, Burgess+18}. This is due to the different
analysis methods chosen. Nonetheless, this observational fact,
combined with the fact that the photospheric emission is inherent to
GRB fireballs motivated the study of photospheric emission as a
possibly key ingredient in the observed prompt spectra.

Due to the weakness of the observed signal, most of the analysis is done on time integrated signal, as simply not enough photons are observed. However, when analyzing the data of bright GRBs where a time-resolved analysis could be and indeed was done, it was proven that indeed some part (but not all) of
the observed prompt spectra can be well fitted with a thermal (Planck)
spectrum \citep{Ryde04, Ryde05, RP09, McGlynn+09, Larsson+11}. This was
confirmed by several recent observations done with the Fermi satellite
\citep{Ackermann+10, Abdo+09a, Ryde+10, Ryde+11, Guiriec+11,
  Iyyani+13, Guiriec+13}.

\begin{figure}
\includegraphics[width=12cm]{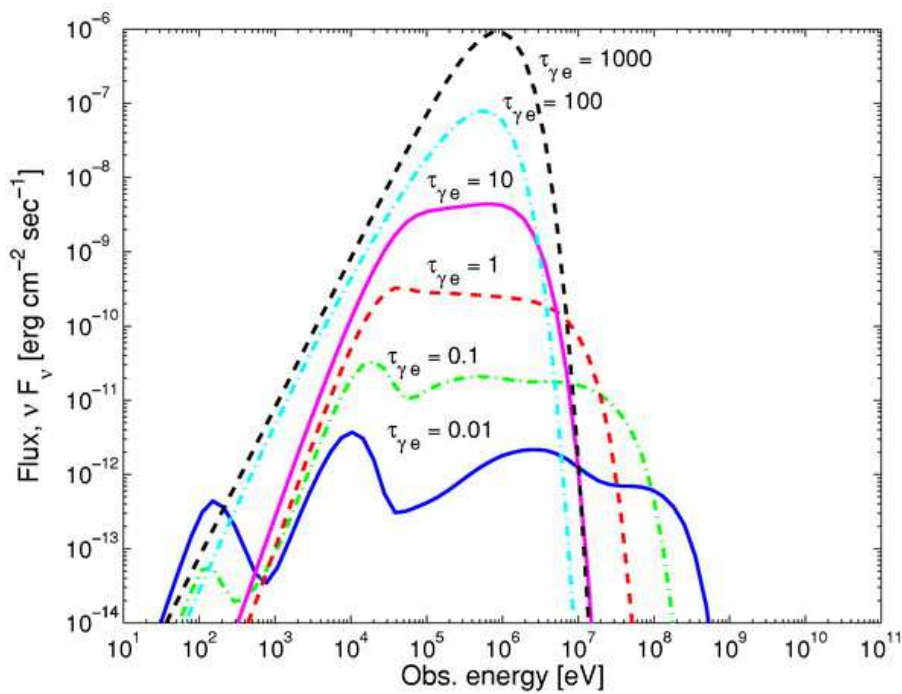}
\caption{Time averaged broad band spectra expected following kinetic
  energy dissipation at various optical depths. For low optical depth,
  the two low energy bumps are due to synchrotron emission and the
  original thermal component, and the high energy bumps are due to
  inverse Compton phenomenon.  At high optical depth, $\tau
  \geq 100$, a Wien peak is formed at $\sim 10 \keV$, and is
  blue-shifted to the MeV range by the bulk Lorentz factor $\simeq
  100$ expected in GRBs.  In the intermediate regime,
$0.1 < \tau < 100$,
a flat energy spectrum above the thermal
  peak is obtained by multiple Compton scattering.
Figure is taken from \citet{PMR06}.
  }
\label{fig:6}
\end{figure}

From a theoretical perspective, photospheric emission that is combined
with other radiative processes was considered as of the early 2000's
\citep{MR00, MRRZ02}. The key issue is that the thermal photons,
similar to the synchrotron photons can be up-scattered by the
energetic electrons. In fact, by definition of the photosphere, they
have to be upscattered as the optical depth below the photosphere is
$> 1$. This implies both modification of the electron distribution
from their initial (accelerated) power law distribution, and
modification of the thermal component itself, in a non-linear way
\citep{PMR05, PMR06}. This naturally leads to a broadening of the
``Planck'' spectrum, which, for a large parameter space region,
resembles the observed one \citep{PMR06}. This is demonstrated in Figure \ref{fig:6}.

This idea of modified Planck spectra gained popularity in recent
years, as it is capable of capturing the key observational GRB
properties in the framework of both photon-dominated and
magnetic-dominated flows \citep{Ioka+07, TMR07, Lazzati+09, LB10,
  Beloborodov10, Mizuta+11, LMB11, Toma+11, Bromberg+11, Levinson12,
  Veres+12, VLP13, Beloborodov13, Hascoet+13, Lazzati+13, AM13, DZ14,
  C-M+15}.

An underlying assumption here is that a population of energetic
particles {\bf can} exist below the photosphere (or close to it). This is not obvious,
as recent works showed that the structure of shock waves, if exist
below the photosphere (``sub-photospheric shocks'') does not enable
the Fermi acceleration process, at least in its classical form
\citep{Levinson12, Beloborodov17, LBV18, LB18}. Nonetheless, particle
heating can still take place below the photosphere via other
mechanisms. For example, due to turbulence cascade
which passes kinetic fluid energy to photons through scattering
\citep{Zrake+18}. Thus, overall, the question of particle heating
below the photosphere and sub-photospheric dissipation is still an
open one.

\begin{figure}
\includegraphics[width=10cm]{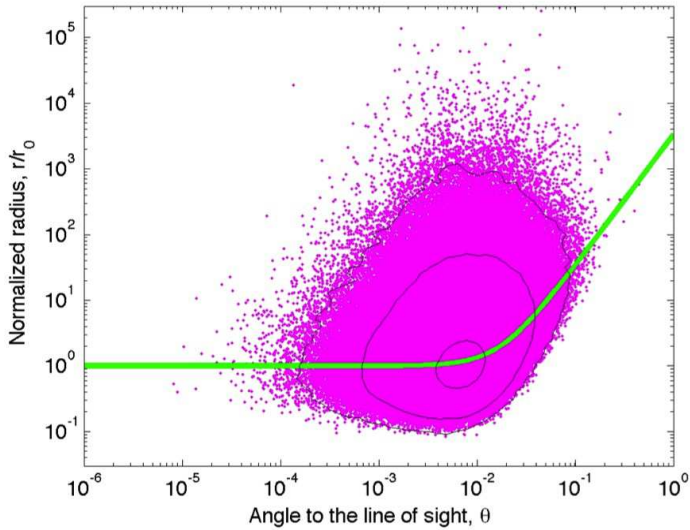}
\caption{The green line represent the (normalized) photospheric radius
  $r_{ph}$ as a function of the angle to the line of sight, $\theta$,
  for spherical explosion. The purple
  dots represent the last scattering locations of photons emitted in
  the center of a relativistic expanding ``fireball'' (using a
  Monte-Carlo simulation). The black lines show contours. Clearly,
  photons can undergo their last scattering at a range of radii and
  angles, leading to the concept of ``vague photosphere''. The
  observed photospheric signal is therefore smeared both in time and
  energy.  Figure taken from \cite{Peer08}.  }
\label{fig:7}
\end{figure}

\begin{figure}
\includegraphics[width=8cm]{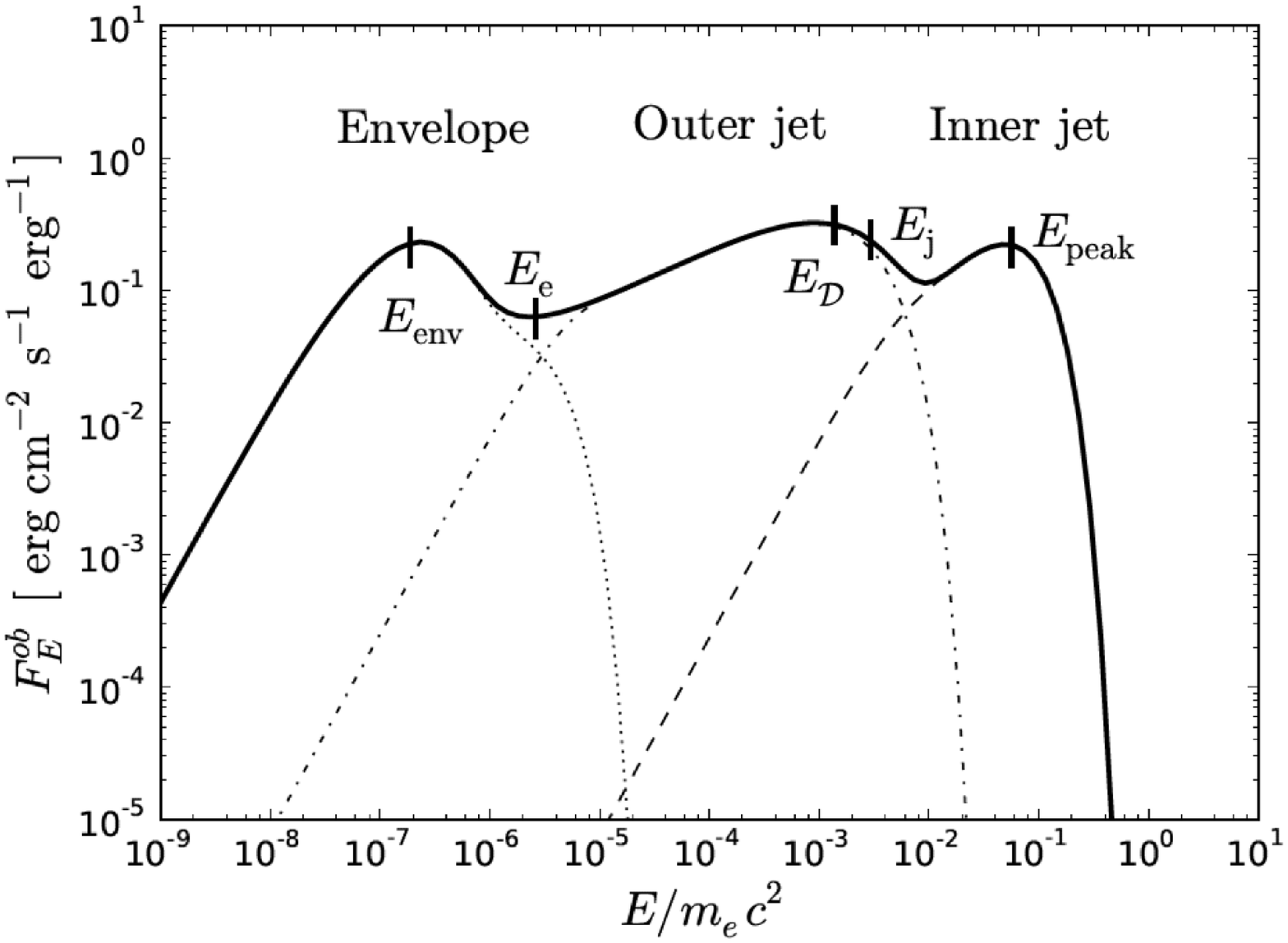}
\includegraphics[width=4cm]{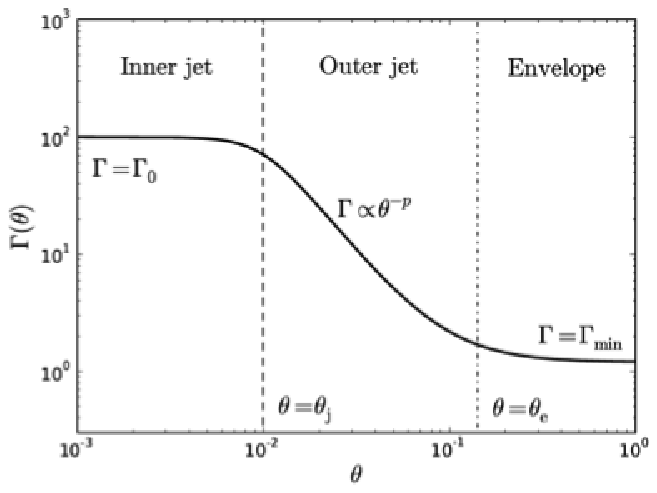}
\caption{{\bf Left.} The observed spectrum that emerges from the optically thick regions of an expanding,
  relativistic jet having a spatial profile, $\Gamma = \Gamma(\theta)$ does
  not resemble the naively expected ``Planck'' spectrum. Separate
  integration of the contributions from the inner jet (where $\Gamma
  \approx \Gamma_0$), outer jet (where $\Gamma$ drops with angle) and
  envelope is shown with dashed, dot dashed and dotted lines,
  respectively. {\bf Right.} The assumed jet profile. Figure taken
  from \citet{LPR13}.  }
\label{fig:8}
\end{figure}

A second, independent way of broadening the ``Planck'' spectra that
enables it to resemble the observed prompt emission spectra of many
GRBs is the relativistic ``limb darkening'' effect, which is geometric
in nature. By definition, the photosphere is a region in space in
which the optical depth to scattering is $> 1$. In a relativistically
expanding plasmas, this surface has a non-trivial shape
\citep{Peer08}.  Furthermore, as photon scattering is probabilistic in
nature, the photospheric region is in a very basic sense ``vague'' -
photons have a finite probability of being scattered anywhere in space
in which particles exist \citep{Peer08, Beloborodov10, Beloborodov11,
  PR11, LPR13, Ruffini+13, Aksenov+13, Ito+13, Vereschagin14}; see Figure \ref{fig:7}. The
exact shape of this ``vague'' photosphere depends on the jet geometry,
and in particular on the jet velocity profile, namely $\Gamma =
\Gamma(r,\theta)$. Under plausible assumptions, this relativistic limb
darkening effect can lead to an observed spectra that does not
resemble at all a ``Planck'' spectra, and, in addition, can be very
highly polarized - up to 40\%, if viewed off the jet axis
\citep{LPR13, LPR14, Chang+14}. This is demonstrated in Figure \ref{fig:8}, taken from \citet{LPR13}.

Identification of this thermal emission component have several
important implications in understanding the conditions inside the
plasma. First, it can be used to directly probe the velocity at the
photospheric radius - the inner most region where any electromagnetic
signal can reach the observer \citep{Peer+07, Larsson+15, Peer+15,
  Wang+17}. Second, identification of a photospheric component can be
used to constrain the magnetization of the outflow \citep{ZM02, DM02,
  ZP09, BP14, BP15}. In a highly magnetized outflow, the photospheric
component is suppressed, and therefore identifying it can be used to
set upper limits on the magnetization.  Furthermore, within the
context of the ``striped wind'' model, the existence of strong photon
field modifies the rate of reconnection \citep{BPL17}.  These
identifications led to the suggestion that possibly the GRB outflow is
initially strongly magnetized (as is suggested within the \citet{BZ77}
mechanism), but that the magnetic field quickly dissipates below the
photosphere \citep{BP14, Meng+18}.

\section{GRB Environments and GRB170817a}
\label{sec:5}

One of the key open questions that had been the subject of extensive
research over the years is the nature of the GRB progenitors. There
are two leading models. The first is the ``collapsar'' model mentioned
above, which involves a core collapse of a massive star, accompanied
by accretion into a black hole \citep[][and references
  therein]{Woosley93, Pac98, Pac98b, Fryer+99, MW99, Popham+99, MacFadyen+01,
  WB06, Sobacchi+17}. The second scenario is the merger of two neutron stars
(NS-NS), or a black hole and a neutron star (BH-NS). The occurrence
rate, as well as the expected energy released, $\sim G M^2/R \sim
10^{53}$~erg (using $M \sim M_\odot$ and $R \gsim R_{sch.}$, the
Schwarzschild radius of stellar-size black hole), are sufficient for
extra-galactic GRBs \citep{Eichler+89, Pac90, Narayan+91, MR92,
  Narayan+92}

The association of long GRBs with core collapse supernova, of type
Ib/c \citep{Galama+98, Hjorth+03, Stanek+03, Campana+06, Pian+06,
  Cobb+10, Starling+11} serves as a ``smoking gun'' to confirm that
indeed, the long GRB population originates from the
``collapsar'' scenario. Indeed, in all cases but two (GRB060505 and
GRB060614) whenever the GRBs were close enough that evidence for
supernovae could be detected, they were indeed observed
\citep{Cano+17}.

While long being suspected, until last year there were only indirect
evidence that short GRBs may be associated with the merger
scenario. These were mainly based on morphologies of the host galaxies
(short GRBs are associated with elliptical galaxies, while long GRBs
reside in younger, star-forming galaxies), as well as their position
in the sky relative to their host galaxy \citep{Gehrels+05,
  Fruchter+06}; \citep[For reviews see, e.g.,][]{Nakar07, GRF09,
  Berger14}. This situation changed with the discovery of the
gravitational wave associated with the short GRB170817
\citep{Abbott+17a, Abbott+17b, Goldstein+17}. This discovery proved
that neutron star - neutron star (NS-NS) merger do indeed produce a
short GRB, thereby providing the missing ``smoking gun''. This event,
though, was unique in many ways - e.g., the large viewing angle
\citep{Alexander+17}, and thus it is not clear whether it is
representative of the entire short GRB population. Indeed, a detailed
analysis show that the environments of short GRBs do not easily fit
this "merger" scenario model \citep{NFP09}.

Despite these uncertainties, it is widely believed that these two
types of progenitors may end up with very different environments. The
merger of binary stars is expected to occur very far from their
birthplace, in an environment whose density is roughly constant, and
equals to the interstellar medium (ISM) density. On the other hand, a
massive star (e.g., a Wolf-Rayet type) is likely to emit strong wind
prior to its collapse \citep{Weaver+77}, resulting in a ``wind'' like
environment, whose density (for a constant mass ejection rate and
constant wind velocity) may vary as $\rho \propto r^{-2}$. I should
stress though that this is a very heuristic picture, as the properties
of the wind emitted by stars in the last episode before they collapse
is highly speculative. Furthermore, even if this is the case, one can
predict the existence of a small ``bump'' in the lightcurve, resulting
from interaction of the GRB blast wave with the wind termination shock
\citep{PW06}. Though this could be very weak
\citep{VanEerten+09}. Indeed, no clear evidence for such a jump (whose
properties are very uncertain) currently exists.

Nonetheless, as early as a few years after the detection of the first
long GRB optical afterglow \citep{VanPar+97} a split between ISM-like
and wind-like environments was observed, with up to 50\% of bursts
found to be consistent with a homogeneous medium  \citep[e.g.,][]{CL99,
  PK01, PK02}.  In later studies, \citep[e.g.,][]{Starling+08,
  Curran+09} ISM-like environments continued to be found in long GRB
afterglows. Further measurements of the spectral and temporal indices
for optical \citep{Oates+12} and X-ray \citep{Oates+15, Li+15,
  Racusin+16, GFP18} afterglows of long GRBs all point to a split in
environment types between wind and ISM.

The theoretical analyses that lead to this conclusion is relatively
simple, as these are based on measurements of the properties of the late
time afterglow. During this stage the outflow is expected to evolve in
a self-similar way. Despite the uncertainty in the detailed of the
processes, it is expected that the velocity profile, the particle
acceleration and the magnetic field generation all follow well defined
scaling laws. These enable the use of the relatively well sampled
afterglow data to infer the properties of the environment at late
times. From this knowledge, one can hope to constrain the nature of
the progenitor, hence the properties of the GRB plasma. The inconsistencies frequently found between the afterglow data of both the long and short GRBs and the simplified environmental models implies that we still have a way to go before understanding the nature of the progenitors, hence the conditions inside the GRB plasmas.

\section{Summary}
\label{sec:summary}

GRBs serve as unique laboratories to many plasma physics
effects. In fact, GRB observations triggered many basic studies in
plasma physics, whose consequences reach far beyond the field of GRBs,
and even extend beyond the realm of astrophysics.

GRBs are the only objects known to produce ultra-relativistic shock
waves, whose Lorentz factors exceed $\Gamma \gsim 100$. As such, they
are the only objects that serve as laboratories to study the properties
of ultra-relativistic shocks. In section \ref{sec:2} I highlight a key
property that directly follow GRB afterglow observations, that of
particle acceleration to high energies in relativistic shock
waves. While the existence of cosmic rays imply that such mechanism
exist, and although the mechanism by which (non-relativistic) shock
waves accelerate particles was discussed in the 1950's by Fermi, the
details of the process in two important limits: (1) the relativistic
limit, and (2) the ``back reaction'' of the accelerated particles on
the shock structure (i.e., the opposite of the ``test particle''
limit) are not known. However, from observations in GRB afterglows, it
is clear that these two limits are the ones that exist in nature.

Section \ref{sec:3} was devoted to discussing magnetic fields in
GRBs. This section was divided into three parts. First, I discussed
the current state of knowledge about the generation of strong magnetic
fields in shock waves. This again is directly motivated from GRB
afterglow observations, which show the existence of strong magnetic
fields during this phase. As these fields are several orders of
magnitude stronger than the compressed magnetic fields in the ISM, they must be
generated by the shock wave itself. It is clear today that the process
of magnetic field generation is intimately connected to the process of
particle acceleration.

I then discussed energy transfer from magnetic fields by the magnetic
reconnection process. This is important in one class of models - the
``Poynting flux'' dominated models, that assume that early on the main
source of energy is magnetic energy. This thus motivates a detailed
study of the reconnection process, as a way of transferring this
energy to the plasma - both as a way of generating the relativistic
jets (accelerating the bulk of the plasma), and as a way of
accelerating individual particles to high energies, giving them a
non-thermal distribution. This last subject was treated separately in
section \ref{sec:3.3}.

In section \ref{sec:4} I discussed the last ingredient of GRB plasmas,
which is the photon field and its interaction with the particle and
magnetic fields. The discussion in this section was divided into two
parts. I first highlighted the ``traditional'' radiative processes
such as synchrotron emission and Compton scattering that are expected
when a population of high energy particles resides in a strongly
magnetized region. As far as we know, these are the conditions that
exist during the (late time) GRB ``afterglow'' phase. I then discussed
the role of the photosphere in section \ref{sec:phot_emission}. The
photosphere exists in the early stages of GRB evolution, and may be
an important ingredient that shapes the prompt emission signal.

However, in addition to shaping the observed prompt emission spectra,
the photosphere affects other aspects of the problem as well. As by
definition, the optical depth to scattering below the photosphere is
$> 1$, there is a strong coupling between the photon and particle
fields. This leads to various effects that modify the structure of
sub-photospheric shock waves, affect the dynamics, and can affect the
magnetic field - particle interactions, via modification of the magnetic
reconnection process. Most importantly, there are several
observational consequences that can be tested.

Finally, I briefly discussed in section \ref{sec:5} our knowledge
about the different environments in GRBs. The current picture is very
puzzling, and there is no simple way to characterize the
environment. The importance of this study lies in the fact that
understanding the environment can provide very important clues about
the nature of the progenitors, hence on the physical conditions inside
the GRB plasmas. Such clues are very difficult to be obtained in any
other way.

Thus, overall, GRBs, being the most extreme objects known, provide a
unique laboratory to study  plasma physics in a unique, relativistic
astrophysical environment. While GRB studies triggered and stimulated
many plasma physics studies in the laboratory, clearly, unfortunately
we cannot mimic the conditions that exist within the GRB environment
in the lab. Thus, in the foreseen future, by large, we will have to
continue rely on GRB observations to provide the necessary input to
test the theories.

As I demonstrated here, as of today there is no consensus on many
basic phenomena which are at the forefront of research. However, as
the study of GRBs is a very active field - both observationally and
theoretically, one can clearly expect a continuous stream of data and
ideas, that will continue to change this field.

\vspace{6pt}

\acknowledgments{AP acknowledges support by the European Research Council via the ERC consolidating grant \#773062 (acronym O.M.J.). I wish to thank Antoine Bret for useful discussions. }

\conflictsofinterest{The authors declare no conflict of interest.}




	\bibliographystyle{aps-nameyear}      

\end{document}